\documentclass[aps,prc,twocolumn,showpacs,amsmath,amssymb,nofootinbib,10pt]{revtex4-2}
\usepackage{graphicx}
\usepackage{amsfonts, amsmath, amssymb}
\usepackage{color}
\usepackage{times}
\usepackage{inputenc}
\usepackage{bm}
\usepackage{ulem}
\usepackage{multirow}
\usepackage{url}
\usepackage{natbib}
\usepackage{mathrsfs}
\usepackage{physics}
\usepackage{comment}
\usepackage[table,xcdraw]{xcolor}
\usepackage{comment}
\usepackage{booktabs}
\usepackage[caption=false]{subfig}
\usepackage[colorlinks=true,citecolor=blue,urlcolor=blue,linkcolor=blue]{hyperref}
\usepackage{wasysym}
\begin{document}
	\title{Influence of dark matter on hybrid and twin stars}
	\author{H. C. Das}
	\affiliation{Institut f\"ur Theoretische Physik, Goethe Universit\"at, Max-von-Laue-Stra\ss{}e~1, D-60438 Frankfurt am Main, Germany}
	\affiliation{INFN Sezione di Catania, Dipartimento di Fisica,
		Via S. Sofia 64, 95123 Catania, Italy}
	\date{\today}
	\begin{abstract}
		We investigate the impact of dark matter (DM) on hybrid and twin stars within a two-fluid framework, where DM and normal matter interact only through gravity. A self-interacting fermionic DM model is considered, while for the nuclear and quark matter we employ relativistic mean-field models and the constant sound-speed parameterization, respectively. Solving the two-fluid Tolman-Oppenheimer-Volkoff equations over a five-dimensional parameter space spanning the quark matter transition pressure $p_t$, energy density jump 
		$\Delta\epsilon$, sound speed $C_s^2$, DM particle mass $m_\chi$, and DM fraction $f_\chi$, we classify the resulting configurations into hybrid stars and four twin star categories and map the onset properties of the quark phase. The effect of DM on the hybrid and twin star population is regime dependent: the DM particle mass determines whether DM forms an extended halo or a compact core, while the DM fraction sets the magnitude of the effect. In the halo regime, DM enhances quark matter formation and increases the number of hybrid and twin stars; in the core regime, it suppresses them. DM also stabilizes hybrid and twin configurations that are unstable in the pure-baryonic case and systematically pushes the nucleon-to-quark transition to higher densities, with the onset pressure, mass, and radius all rising with the DM fraction, potentially hiding quark matter from current neutron star observations.
	\end{abstract}
	\maketitle
	\section{Introduction}
	\label{sec:intro}
	A wide range of astrophysical and cosmological observations such as galactic rotation curves, gravitational lensing, the cosmic microwave background, and the growth of large-scale structure, firmly establishes that most of the matter content of the Universe is dark. Despite this overwhelming dynamical evidence, the particle nature of dark matter (DM), and in particular its mass and interaction strength, remains unknown. Identifying these properties is one of the central open problems of contemporary physics and is pursued through complementary strategies: direct detection in underground laboratories, indirect detection of annihilation or decay products, collider searches, and astrophysical probes. Among the latter, compact stars have emerged as particularly powerful natural laboratories for DM.
	
	Neutron stars (NSs) are especially well suited to this purpose because of their immense gravitational potential and extremely high baryonic density \cite{Bell_2021}. Being roughly twice as massive as the Sun and often residing in DM-rich regions, NSs can capture and accumulate DM efficiently \cite{Goldman_1989, Kouvaris_2008}. The amount of DM accreted depends on three factors: the type and properties of the DM particle, the surrounding astrophysical environment, and the internal composition of the star. Once present, DM can modify observable bulk properties such as the mass, radius, tidal deformability, and surface luminosity, and can alter the thermal evolution of the star \cite{Das_2020, Das_curv_2021, Das_insp_2021, Das_GW19_2021, Das_fmode_2021, Das_Gala_2022, Rutherford_2023, Liu_2024, Avila_2024, Zhou_2025}. NSs therefore provide an indirect but sensitive route to constraining DM, especially when combined with multimessenger data \cite{Rutherford_2023, Rutherford_2025}.
	
	The properties of DM-admixed neutron stars (DMNS) have been widely studied in recent years; comprehensive reviews of the field can be found in Refs.~\cite{Bramante_2023, Grippa_2025}. Tidal deformability, accessible through gravitational-wave observations, has emerged as one of the principal probes. A bosonic DM core stiffens the equation of state (EOS) and raises the tidal parameter, rendering such stars distinguishable from purely hadronic ones~\cite{Karkevandi_2022}; this analysis has since been extended to both fermionic and bosonic candidates over a wide range of particle masses~\cite{Leung_2022}. Combining mass, radius, and tidal measurements jointly places much tighter bounds on the DM parameters than any single observable alone~\cite{Mariani_2024}. Bayesian analyses using precise NICER radius data have yielded meaningful constraints on both bosonic and fermionic asymmetric DM~\cite{Rutherford_2023, Rutherford_2025}, the two DM families leave distinguishable imprints on the NS population~\cite{Arvikar_2026}, and simultaneous mass-radius measurements are sensitive to DM fractions as small as a few percent~\cite{Zhang_2025}.
		
	Beyond static structure, nonradial oscillations provide an independent window into the stellar interior. The fundamental ($f$-) mode frequency, in principle accessible to future gravitational-wave detectors, carries clear imprints of DM in quarkyonic stars admixed with DM~\cite{Dey_2025}. Distinct $f$-mode patterns emerge when hyperons, quark matter, and the sexaquark DM candidate coexist~\cite{Shahrbaf_fmode_2026}, and a fully general-relativistic $f$-mode calculation for gravitationally bound DM has been carried out~\cite{Routaray_2026}. At the level of universal relations, the tight correlation between $f$-mode frequency and stellar compactness is preserved in two-fluid DMNS~\cite{Sotani_2025}, and I-Love-Q universality has been extended to DM-admixed stars, yielding a ``Dark I-Love-Q'' family of relations~\cite{Wu_2025}. DM also influences the thermal and rotational evolution: annihilation of accumulated DM slows the cooling of a newly born NS, potentially leaving a detectable thermal signal~\cite{Issifu_2025}, while an independently rotating DM component modifies the maximum spin rate and moment of inertia~\cite{Cipriani_2025}. When DM coexists with exotic nuclear phases, the structural effects are equally rich. The EOS for the simultaneous presence of hyperons, bosonic DM, and quark matter has been mapped out in Ref.~\cite{Shahrbaf_2026}. DM shifts the nucleon-to-quark transition pressure in hybrid NSs~\cite{Wei_2026}, and the tidal deformabilities of the two twin-star branches can differ by orders of magnitude, rendering them clearly distinguishable in gravitational-wave observations even when their masses nearly coincide~\cite{Lu_2025}.
	
	In this study, we explore DM effects on hybrid/twin stars scenarios in the two-fluid framework~\cite{Liu_2024, Rutherford_2023}, in which DM and baryonic matter interact exclusively through gravity; the DM model (see Sec.~\ref{subsec:DM_EOS}) and its properties are described in detail in Sec.~\ref{sec:2f}. At the high densities reached in NS cores, hadronic matter may undergo deconfinement into quark matter \cite{Prakash_2002, Annala_2020}, giving rise to a distinct family of compact objects known as hybrid stars and, under suitable conditions, to twin stars: pairs of stars with nearly identical masses but appreciably different radii \cite{Zdunik_2013, Chamel_2013, Alford_2013, Christian_2018, Montana_2019}. Because the microscopic quark EOS is still poorly constrained, it is convenient to describe the deconfined phase with the constant-speed-of-sound (CSS) parameterization \cite{Zdunik_2013, Alford_2013, Chamel_2013}, in which the quark sector is characterized by just three quantities as mentioned below. The existence and type of a hybrid or twin configuration are then governed by these parameters together with the Seidov stability criterion \cite{Seidov_1971}.
	
	Recent studies have begun to examine DM together with quark degrees of freedom \cite{Lenzi_2023, Biesdorf_2025}. Using the MIT bag model,  Ref. \cite{Lenzi_2023} found that DM shifts the energy-density discontinuity, reducing the minimum mass of the quark core and modifying the tidal deformability.  Biesdorf {\it et al.}~\cite{Biesdorf_2025} reported that DM raises the central pressure, thereby affecting hybrid star formation, and predicted a new class of objects called ``dark oysters'', in which a large DM halo encloses a small nuclear-matter core. These works, however, were carried out either in a single-fluid setting or for a specific quark model, and they did not systematically address how DM reshapes the {\it population} of hybrid and twin stars across the quark matter parameter space, nor how this depends on whether the DM forms a core or a halo.
	
	This last point defines the central physical question of the present work. Because in the two-fluid picture DM couples only gravitationally, its entire effect on the star is structural: it reshapes the pressure and density profile of the baryonic component. Whether this reshaping {\it favors} or {\it disfavors} the appearance of a stable quark core is not obvious in advance, and it cannot be read off from the model construction. It depends on where the DM resides relative to the baryonic matter, which is itself fixed by the DM particle mass: a light particle is supported by degeneracy pressure over a large volume and forms an extended halo, whereas a heavy particle is confined to a compact central core. We will show that these two scenarios act on the baryonic core in opposite ways; one helping the star to reach the quark phase, the other preventing it from sustaining one, so that the DM particle mass effectively sets the {\it sign} of the DM influence on the hybrid/twin-star population, while the DM fraction controls its magnitude. Establishing this competition, and quantifying it through the onset properties of the quark phase, is the main goal of the paper.
	\begin{table}
		\centering
		\caption{Nuclear matter saturation properties and neutron star observables for NITR-1 \cite{Routaray_2023} and DD2 \cite{Typel_2010} models.}
		\label{tab:nuclear_properties}
		\begin{tabular}{lcccccccc}
			\toprule
			Model & $\rho_{\text{sat}}$ & $\mathcal{E}_{\text{sat}}$ & $K_{\text{sat}}$ & $J_{\text{sat}}$ & $L_{\text{sat}}$ & $M_{\text{max}}$ & $R_{1.4}$ & $\Lambda_{1.4}$ \\
			& (fm$^{-3}$)         & (MeV)                       & (MeV)             & (MeV)             & (MeV)             & ($M_{\odot}$)     & (km)      &                 \\
			\midrule
			NITR-1 & 0.151 & $-16.3$ & 200 & 30.9 & 62 & 2.34 & 12.7 & 527 \\
			DD2    & 0.149 & $-16.0$ & 243 & 31.7 & 55 & 2.42 & 13.2 & 680 \\
			\bottomrule
		\end{tabular}
	\end{table}
	\begin{figure*}
		\centering
		\includegraphics[width=0.5\textwidth]{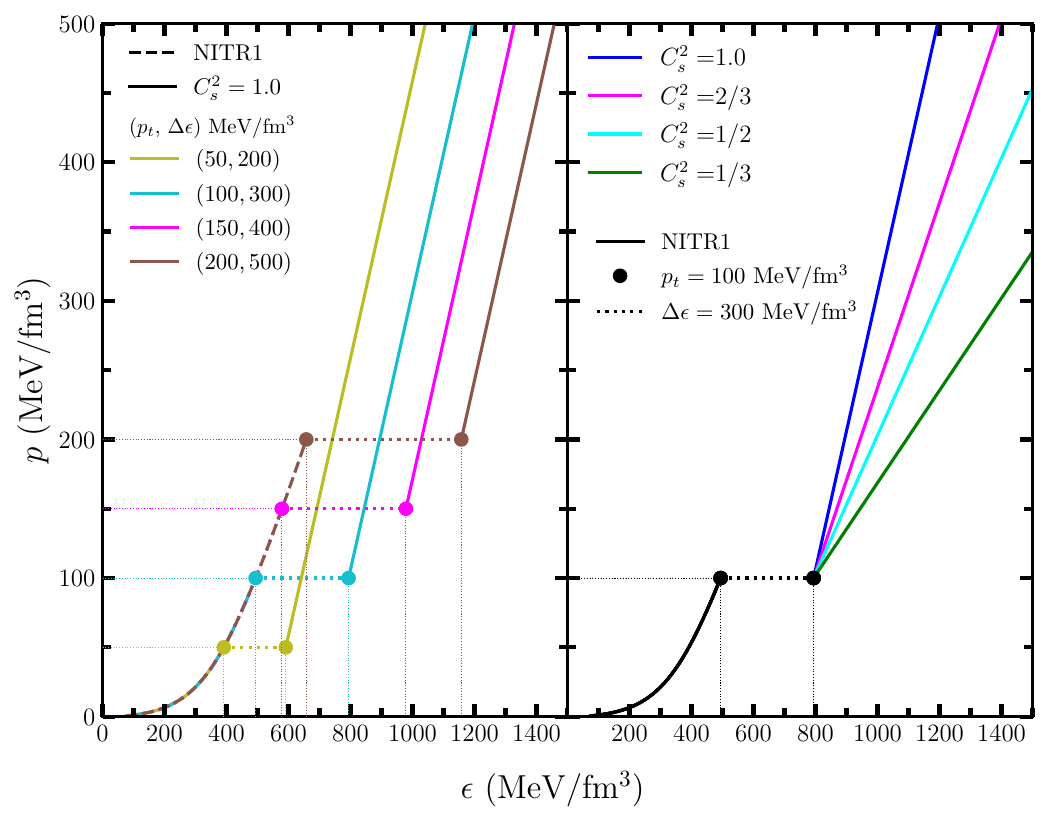}
		\includegraphics[width=0.465\textwidth]{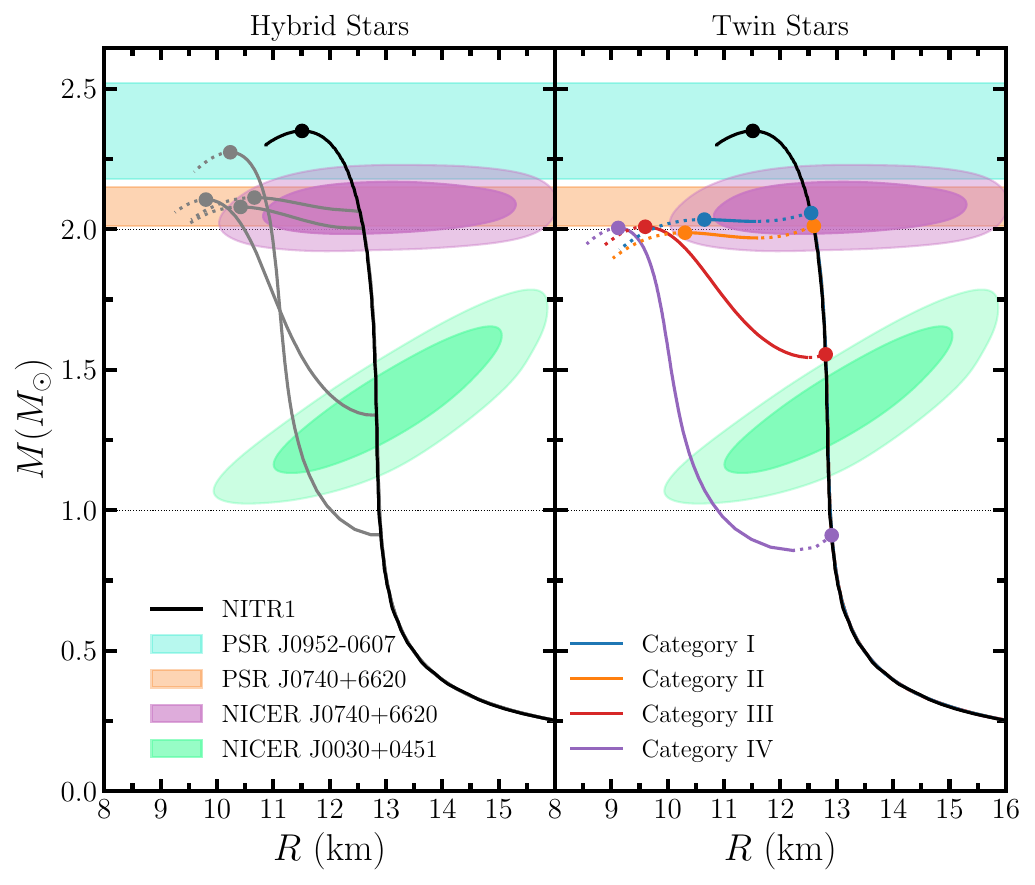}
		\caption{{\it Left:} EOSs are shown for different $p_t$ and $\Delta \epsilon$ values with $C_s^2=1$, and for a fixed $p_t$ and $\Delta \epsilon$ with different values of $C_s^2$. {\it Right:}  hybrid stars and twin stars are shown with different values of $p_t$ and $\Delta \epsilon$ with $C_s^2=1$. The dotted lines represent the unstable region in the mass-radius sequences. $\CIRCLE$ marker represents the maximum mass configuration for the NITR-1 case, and colors represent the same configuration for the primary and secondary branches. The observational constraints, such as maximum mass from two pulsars (PSR J0952-0607 \cite{Romani_2022}, PSR J0740+6620 \cite{Fonseca_2021}) and mass-radius contours from two different NICER measurements (J0740+6620 \cite{Salmi_2024}, J0030+04531 \cite{Vinciguerra_2024}), are also shown.}
		\label{fig:eos_mr}
	\end{figure*}
	
	We model the hadronic sector with two relativistic mean-field EOSs of different stiffness, NITR-1 \cite{Routaray_2024} and DD2 \cite{Typel_2010} (see Table \ref{tab:nuclear_properties} for nuclear matter properties and NS observables), so that the robustness of the conclusions against the nucleonic EOS can be assessed; the quark sector with the CSS parameterization, with $C_s^2$ varied from intermediate soft to stiff values; and the DM sector with a self-interacting fermionic model. Solving the two-fluid TOV equations over this parameter space, we consider two representative DM particle masses chosen to bracket the halo and core regimes, together with three DM mass fractions, classify the resulting configurations into hybrid star and the four twin star categories, count their populations, and map the onset pressure, mass, and radius of the quark phase. We further use the $2\,M_\odot$ constraint on both hybrid star and twin star populations to show how observations restrict the viable parameter space. The paper is organized as follows. Sections~\ref{sec:EOS}--\ref{sec:claasification} describe the EOS construction, the characterization of hybrid and twin stars, and the classification scheme; Sec.~\ref{sec:2f} presents the two-fluid results and the physical discussion; and Sec.~\ref{summary} summarizes our conclusions.
	\section{Equation of states of the system}
	\label{sec:EOS}
	\subsection{EOS for nucleonic and quark matter}
	\label{subsec:NM_EOS}
	The EOS of the NS can be calculated from crust to core based on the types of particles present in each region and interactions between them. The outer crust is composed of the neutron-rich isotopes embedded in the Coulomb lattice in the degenerate electron gas. As a result, the nuclei capture electrons and form different isotopes in the region $26 \leq Z \leq 40$. The degenerate electrons solely determine the pressure in this region. Hence, the EOSs for this region are relatively well known. With a density of more than that, the nuclei are highly neutron-rich, and they do not bind the additional neutrons. Therefore, the neutrons are dripped out from the nuclei. The region defines the boundary between the outer and inner crust. As the densities increase further, the nucleons form clusters in Coulomb crystals of neutron-rich nuclei embedded by uniform electrons. The minimization of the system's energy produces different exotic structures known as pasta \cite{Parmar_2022}. 
	
	Therefore, for the inner-crust part, we adopt the Negele-Vautherin \cite{Negele_1973} EOS within the density range (0.001 fm$^{-3} < \rho < \rho_t$), and for the outer crust ($\rho < 0.001$ fm$^{-3}$) the Baym-Pethick-Sutherland \cite{Baym_1971} EOS. By ensuring a smooth transition in pressure and energy density between the crustal and core EOS branches, a transition density of approximately $\rho_t \approx 0.08$ fm$^{-3}$ is obtained. For nucleonic EOS, we chose ``DD2" \cite{Typel_2010} and our recently developed model ``NITR-1" \cite{Routaray_2024} based on the relativistic mean-field model in the density above $\rho>0.08$ fm$^{-3}$. This EOS satisfies the latest massive pulsar mass alongside recent NICER observations data, see Table \ref{tab:nuclear_properties}. 
	
	At high-density regions inside a star, nucleons break into their constituent quarks. The interactions between quarks differ from those between nucleons due to the fundamental nature of the strong force at the quark level. For quark EOS, two widely used models are employed to investigate quark interactions: (i) the MIT bag model, and (ii) the Nambu-Jona-Lasinio model are proposed in the literature. However, a simpler alternative model, CSS, which effectively mimics quark interactions, was first proposed in \cite{Zdunik_2013, Alford_2013}. In the CSS model, three important parameters, such as the constant speed of sound ($C_s^2 = dp/d\epsilon = {\rm const}$), transition pressure ($p_t$), and energy discontinuity ($\Delta \epsilon$) at the transition density, play a significant role in constructing the EOS \cite{Zdunik_2013, Alford_2013, Chamel_2013}. 
	
	The system's EOS can be written as \cite{Christian_2018, Montana_2019}:
	\begin{align}
		\epsilon(p) = 
		\begin{cases} 
			\epsilon_{\rm n}(p) & \text{for } p < p_t \; , \\
			\epsilon_{\rm n}(p_t) + \Delta \epsilon + (p - p_t)/C_s^2 & \text{for } p > p_t \; ,
		\end{cases}
		\label{eq:EOS}
	\end{align}
	where $\epsilon_{\rm n} (p)$ is the EOS, including crust and nucleonic core as a function of pressure. The $C_s^2$ has value within the range $\frac{1}{3} < C_s^2 < 1$. Perturbative QCD sets the lower bound at $C_s^2 = \frac{1}{3}$, while $C_s^2 = 1$ corresponds to the stiffest EOS permissible under this model. The EOS is depicted in the left panel of Fig.~\ref{fig:eos_mr} for the NITR-1 model as a representative case.          
	
	The value of $p_t$ predicts the appearance of quarks at the transition density. However, $C_s^2$ predicting the nature of the EOS depends on its magnitude, as is clearly shown in the left panel of the figure. Using $C_s^2=1$ in Eq. (\ref{eq:EOS}), it depends only on $p_t$ and $\Delta \epsilon$. These two parameters play a crucial role in deciding the nature of the star, whether twin or hybrid, by solving the Tolman-Oppenheimer-Volkoff (TOV) equations as shown on the right side of Fig. \ref{fig:eos_mr}.
	\begin{figure*}
		\centering
		\includegraphics[width=0.5\textwidth]{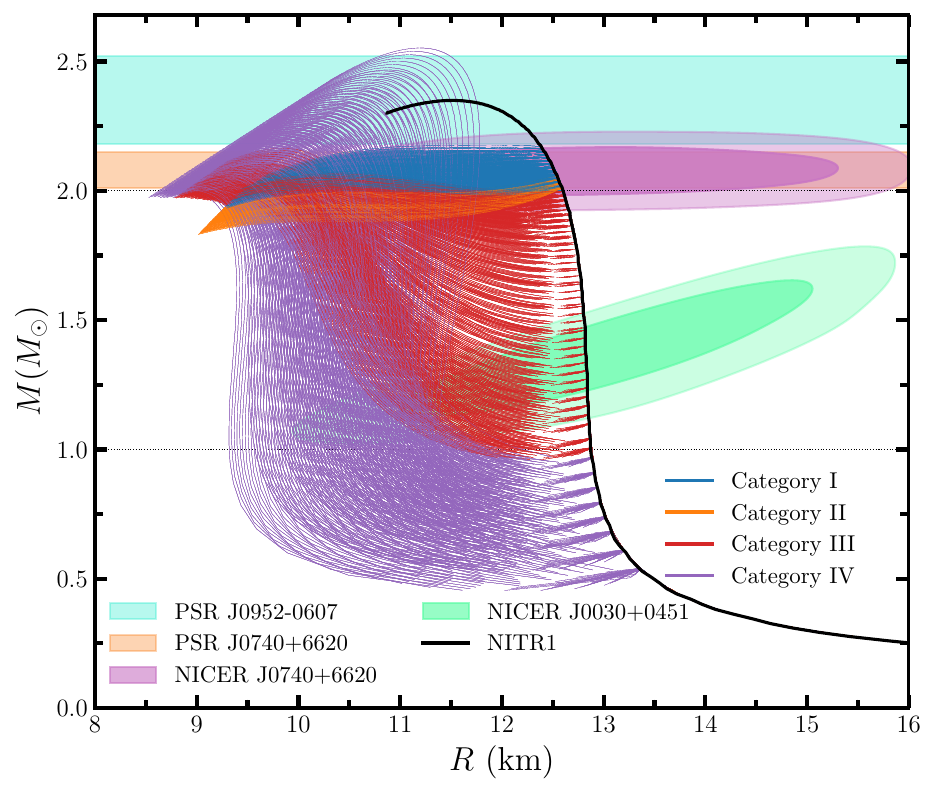}
		\includegraphics[width=0.5\textwidth]{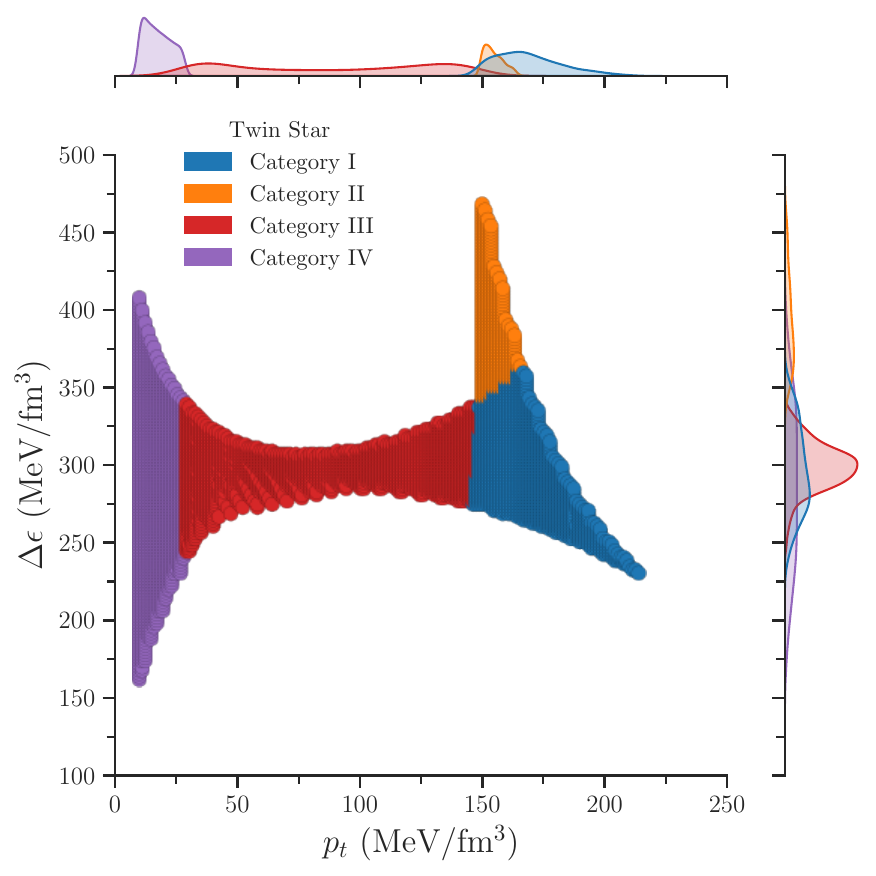}
		\caption{{\it Left:} the mass-radius relations for all categories of twin star are shown only for NITR-1 as a representative case. {\it Right:} the joint plot for the phase space of only twin star characterized by $p_t$ and $\Delta \epsilon$. The upper and side plots are the marginalized PDF for each category. The colors for categorized twin star are same as Fig.~\ref{fig:eos_mr}.}
		\label{fig:joint_mr_twin star}
	\end{figure*}
	\subsection{EOS for dark matter}
	\label{subsec:DM_EOS}
	The DM component is described by a self-interacting fermionic model, in which asymmetric DM particles of mass $m_\chi$ interact via a repulsive vector mediator $\phi_\mu$. Unlike the single-fluid approach, in which DM couples directly to nucleons through a nongravitational channel~\cite{Das_2020}, the two-fluid treatment is the natural choice when the DM--baryon cross section is negligible; it also has a practical advantage: the DM and baryonic sectors are modeled independently and coupled solely through the stellar structure equations, so the influence of DM on the star is purely structural and can be isolated easily. The Lagrangian of the DM sector reads  \cite{Routaray_2025}
	\begin{equation}
		\mathcal{L}_\text{DM} = \bar{\chi}(i\gamma^\mu D_\mu - m_\chi)\chi 
		+ \frac{1}{2}m_\phi^2\phi_\mu\phi^\mu 
		- \frac{1}{4}\Omega_{\mu\nu}\Omega^{\mu\nu} \:,
	\end{equation}
	where $\chi$ and $\phi_\mu$ are the fermionic DM field and the vector boson field with masses $m_\chi$ and $m_\phi$, respectively, $D_\mu = \partial_\mu + ig_\chi\phi_\mu$ is the covariant derivative with coupling constant $g_\chi$, and $\Omega_{\mu\nu} = \partial_\mu\phi_\nu - \partial_\nu\phi_\mu$	is the field strength tensor. The resulting DM energy density and pressure are given by~\cite{Narain_2006,Liu_2024}
	\begin{align}
		\varepsilon_\chi &= \frac{m_\chi^4}{8\pi^2}
		\Big[x\sqrt{1+x^2}(2x^2+1) - \sinh^{-1}(x)\Big] + \left(\frac{y n_\chi}{m_\chi}\right)^2 \:,
		\\
		p_\chi &= \frac{m_\chi^4}{8\pi^2}
		\Big[x\sqrt{1+x^2}\!\left(\frac{2x^2}{3}-1\right) 
		+ \sinh^{-1}(x)\Big] +  \left(\frac{y n_\chi}{m_\chi}\right)^2 \:,
	\end{align}
	where
	\begin{equation}
		x = \frac{k_\chi}{m_\chi} = \frac{(3\pi^2 n_\chi)^{1/3}}{m_\chi}
	\end{equation}
	is the dimensionless Fermi momentum parameter, with $k_\chi$ and $n_\chi$ being the DM Fermi momentum and number density, respectively. $y$ is the self-interaction parameter related as $y \equiv g_\chi m_\chi/(\sqrt{2}\,m_\phi)$.
	
	The parameters $m_\chi$ and $y$ are not independent. Observations of colliding galaxy clusters impose constraints on the DM self-interaction cross section, requiring 
	$\sigma/m_\chi \sim 0.1$--$10\ \mathrm{cm^2\,g^{-1}}$~\cite{Markevitch_2004,Kaplinghat_2016,Sagunski_2021}. Adopting the Born approximation, which remains accurate for $m_\chi \lesssim 1$~GeV, and fixing $\sigma/m_\chi = 1\ \mathrm{cm^2\,g^{-1}}$, one obtains \cite{Liu_2024}
	\begin{align}
		y^4 &= \pi\, m_\chi^3\, (\sigma/m_\chi) \:, \nonumber \\
		y   &\approx 10.94\, m_1^{3/4} \:,
		\label{eq:y_constraint}
	\end{align}
	where $m_1 \equiv m_\chi/1\,\mathrm{GeV}$. With this constraint applied, the DM EOS depends on a single free parameter $m_\chi$, which simultaneously controls the spatial extent of the DM distribution inside the star: a lighter particle spreads into an extended halo ($R_\chi > R_N$), while a heavier particle settles into a compact core ($R_\chi < R_N$), as we discuss in detail in Sec.~\ref{sec:2f}.
	
	For completeness, we note that alternative DM models have also been explored in the literature: bosonic candidates (scalar fields, axion-like particles, ultralight fields) that may undergo Bose-Einstein condensation in the NS core, and Higgs-portal or vector-portal models with nongravitational DM--baryon couplings~\cite{Grippa_2025,Karkevandi_2022,Dey_2025,Hajkarim_2025,Thakur_2026}. While quantitative results depend on the DM model chosen, the present work focuses on the self-interacting fermionic description, and the robustness of our conclusions across different DM models is discussed in Sec.~\ref{summary}.
	\section{Characterization of Hybrid/Twin Stars}
	\label{sec:charc}
	Due to the onset of a first-order phase transition from nucleonic matter to quark degrees of freedom, a new stable branch emerges, known as hybrid star. Twin star emerges from hybrid star with a disconnected feature. The nature of these configurations is solely characterized by two key parameters $p_t$ and $\Delta\epsilon$ at the phase transition for a fixed value of $C_s^2$. The stability condition for both the hadronic and hybrid star sequences must be satisfied:
	\begin{align}
		\frac{\partial M}{\partial \epsilon_c}>0 \; ,
	\end{align}
	
	where $M$ and $\epsilon_c$ are the mass and central energy density of the star in the sequence. This condition ensures that each configuration on the mass-radius curve corresponds to a stable equilibrium solution. Typical scenarios are depicted in Fig. \ref{fig:eos_mr}, both for hybrid and twin star configurations by choosing different values of $p_t$ and $\Delta \epsilon$ for $C_s^2=1$. The observational constraints are also shown to examine the consistency of model parameters. For a fixed value of $p_t$ and $\Delta\epsilon$, the resulting configuration can correspond to either a hybrid star (connected to the main branch) or a twin star (disconnected from the main branch).
	
	It has become evident that when the central energy density ($\epsilon_c$) reaches the transition energy density, $\epsilon_t$, two possible outcomes occur: either the star becomes unstable or it remains stable even after the formation of a quark core, and it is solely decided by $\Delta \epsilon$. This behavior is governed by the Seidov criterion, which defines a critical threshold for the energy density discontinuity, expressed as \cite{Seidov_1971}:
	\begin{align}
		\Delta \epsilon_{\rm crit} = \frac{1}{2}\epsilon_t + \frac{3}{2} p_t\; ,
	\end{align}
	where $\Delta\epsilon_{\rm crit}$ is the critical energy density discontinuity below which a stable hybrid star is connected to the hadronic branch. When the Seidov limit is reached, the sequence of stars becomes unstable immediately; however, the stability of stars could still be possible by choosing certain values of $p_t$ and 
	$\Delta\epsilon$. 
	\section{Classification Criteria}
	\label{sec:claasification}
	The classification is based on the maximum mass values \( m_{1,\text{max}} \) and \( m_{2,\text{max}} \) as follows \cite{Christian_2018, Montana_2019}:
	\begin{itemize}
		\item \textbf{Category I:} \( m_{1,\text{max}} > 2.0 \) and \( m_{2,\text{max}} > 2.0 \),
		\item \textbf{Category II:} \( m_{1,\text{max}} \geq 2.0 \) and \( m_{2,\text{max}} < 2.0 \),
		\item \textbf{Category III:} \( 1.0 \leq m_{1,\text{max}} \leq 2.0 \) and \( m_{2,\text{max}} \geq 2.0 \),
		\item \textbf{Category IV:} \( m_{1,\text{max}} \leq 1.0 \) and \( m_{2,\text{max}} \geq 2.0 \).
	\end{itemize}
	Here, \( m_{1,\text{max}} \) and \( m_{2,\text{max}} \) denote the maximum masses of the primary and secondary branches, respectively. If none of the above conditions are satisfied, the configuration is categorized as ``uncategorized twin star,'' which is less significant since the maximum masses (either primary or secondary) do not satisfy the $2M_\odot$ constraint. All categories of twin and hybrid star are illustrated in Fig.~\ref{fig:eos_mr}, highlighting their dependence on the transition pressure \( p_t \) and the discontinuity in energy density \( \Delta \epsilon \). 
	
	As discussed earlier, the stellar characteristics depend mainly on three parameters: the transition pressure \( p_t \), the energy-density discontinuity \( \Delta \epsilon \), and the fixed sound speed \( C_s^2 \). We therefore explore the full parameter space of \( p_t \) and \( \Delta \epsilon \) with the following ranges, chosen in accordance with previous studies \cite{Christian_2018, Montana_2019}:
	\begin{align}
		\Delta \epsilon &= (100, 500, N) \, \text{MeV/fm$^3$}, \nonumber \\
		p_t &= (10, 250, N) \, \text{MeV/fm$^3$}, 
		\label{eq:grid}
	\end{align}
	where \( N = 200 \), yielding 40,000 grid points. The resulting phase space for twin star, along with marginalized probability distribution functions (PDFs), is shown in Fig.~\ref{fig:joint_mr_twin star} for $C_s^2=1$ as a representative case.
	
	We find that Category III spans a wide range of \( p_t \) values within a relatively narrow range of \( \Delta \epsilon \), consistent with earlier studies \cite{Christian_2018}. Conversely, Category IV exhibits the opposite behavior, with a broad range of \( \Delta \epsilon \) and a narrow range of \( p_t \). Categories I and II occupy intermediate regions, shifted toward higher values of both \( p_t \) and \( \Delta \epsilon \). Each category is separated by approximate boundaries defined by the minimum and maximum values of \( p_t \) and \( \Delta \epsilon \), implying that unique parameter combinations determine the nature of twin star. The peaks of the marginalized PDFs further indicate the most probable values of \( p_t \) and \( \Delta \epsilon \) for each category.  Establishing the twin star phase space in the pure-baryonic case is essential; once this baseline is known, the shifts induced by DM admixture or other exotic degrees of freedom become immediately identifiable and physically interpretable.
	\begin{figure*}
		\centering
		\includegraphics[width=1.02\textwidth]{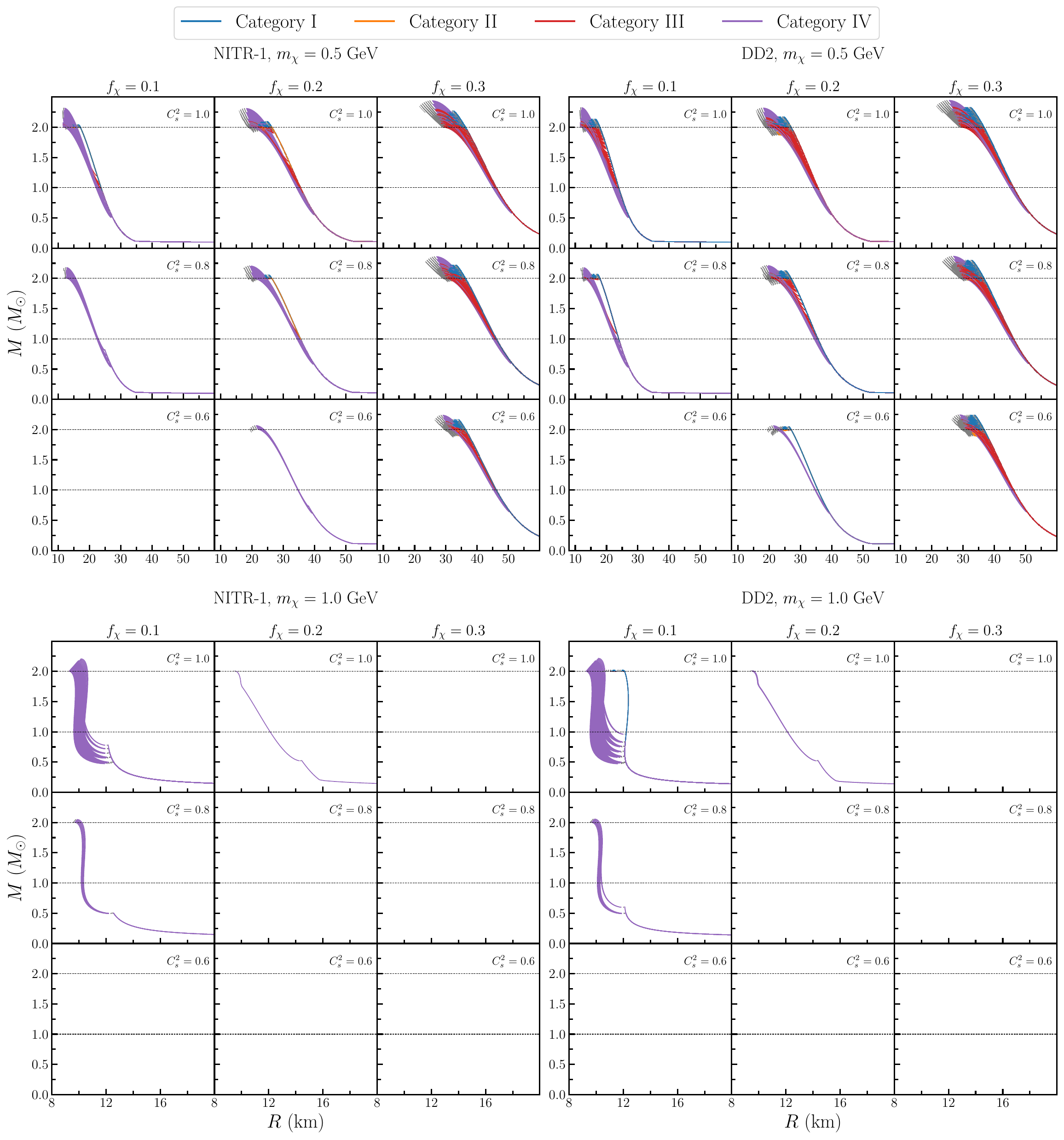}
		\caption{$M-R$ relations for twin star are shown for both NITR-1 and DD2 with $C_s^2 =0.6, 0.8,$ and 1.0 with three DMF having $m_\chi=0.5$ GeV (above) and 1.0 GeV (below). The colors for categorized twin star are same as in Fig.~\ref{fig:eos_mr}.}
		\label{fig:mr_2f_twin}
	\end{figure*}
	\section{Two-Fluid System}
	\label{sec:2f}
In a two-fluid system, nuclear matter (nucleons and quarks) acts as the primary fluid and DM as the secondary, with the two fluids interacting only through gravity, requiring the two-fluid TOV equations to be solved~\cite{Liu_2024}. The EOSs of both fluids are the primary input to solve the TOV equations, as described in Sec.~\ref{sec:EOS}. In this case, the system is governed by five parameters: the quark matter parameters ($p_t$, $\Delta\epsilon$, and $C_s^2$), the DM parameters ($m_\chi$ and $f_\chi = M_\chi/M$), for a fixed nuclear matter EOS. A full uniform scan of this five-dimensional space is computationally expensive, so our sampling is guided by the physics rather than brute force. For the quark sector, we scan $p_t$ and $\Delta\epsilon$ on one-fourth of the grid of Eq.~(\ref{eq:grid}) and consider three values $C_s^2 = 0.6$, $0.8$, and $1.0$, covering the range from a relatively soft to the stiffest quark phase allowed by the model. Although the lowest physically motivated value is $C_s^2 = 1/3$, we find that the number of hybrid star/twin star configurations at this value is negligibly small, and we therefore take $C_s^2 = 0.6$ as our lower bound.

	For the DM sector we choose two particle masses, $m_\chi = 0.5$ and $1.0$~GeV,  and three fractions, $f_\chi = 0.1$, $0.2$, and $0.3$. The two masses are not  arbitrary. In our earlier works~\cite{Routaray_2023, Liu_2024}, we found that a lighter DM  particle ($m_\chi \lesssim 0.5$~GeV) forms an extended halo around the star  ($R_\chi > R_N$), while a heavier one ($m_\chi \gtrsim 1.0$~GeV) settles into a compact core ($R_\chi < R_N$). The values $0.5$ and $1.0$~GeV therefore represent two qualitatively different configurations, letting us compare the halo and core regimes directly within a manageable computational budget. The three DM fractions cover the range typically considered observationally relevant 
	for neutron stars. Since $m_\chi$ controls whether a halo or core forms and $f_\chi$ controls how strongly it affects the stellar structure, this parameter choice is enough to show how the DM configurations and its amount separately influence the star's ability to reach and sustain a quark core. A finer grid in $m_\chi$ and $f_\chi$, including masses near the halo-core boundary, would sharpen the quantitative boundaries found below but is not expected to change the qualitative picture.
	
	Before turning to the results, it is worth understanding how the DM particle mass controls the spatial distribution of DM inside the star, since this single property drives every trend reported below. DM is modeled as a degenerate Fermi gas, and the size of a self-gravitating Fermi gas is determined by the balance between gravity and the degeneracy pressure of its constituent particles. For a lighter particle, the degeneracy pressure supports the DM over a much larger volume, so it spreads into a wide, low-density cloud. For a heavier particle, the same amount of DM is packed into a small, dense region, and both the characteristic size and the maximum mass of the Fermi gas decrease as the particle mass increases.
	
	The DM particle mass therefore acts as a switch between two very different configurations. For $m_\chi = 0.5$~GeV, the DM extends beyond the baryonic surface and forms a diffuse halo surrounding the visible star. For $m_\chi = 1.0$~GeV, the DM shrinks below the baryonic surface and settles at the center as a compact core. These two configurations act on the baryonic matter in opposite ways, and this is the origin of the competition studied in this work. A DM halo adds gravitational binding from outside while leaving the baryonic core only mildly compressed, gently raising the central density and helping the star reach the hadron-quark transition. A DM core, on the other hand, sits directly at the center, strongly compresses the baryonic matter, and drives the configuration toward gravitational instability at a lower central density, so the star can no longer sustain a stable quark core. With this picture in mind, we now examine the twin and hybrid star results are shown below.
	\begin{figure*}
		\centering
		\includegraphics[width=1.02\textwidth]{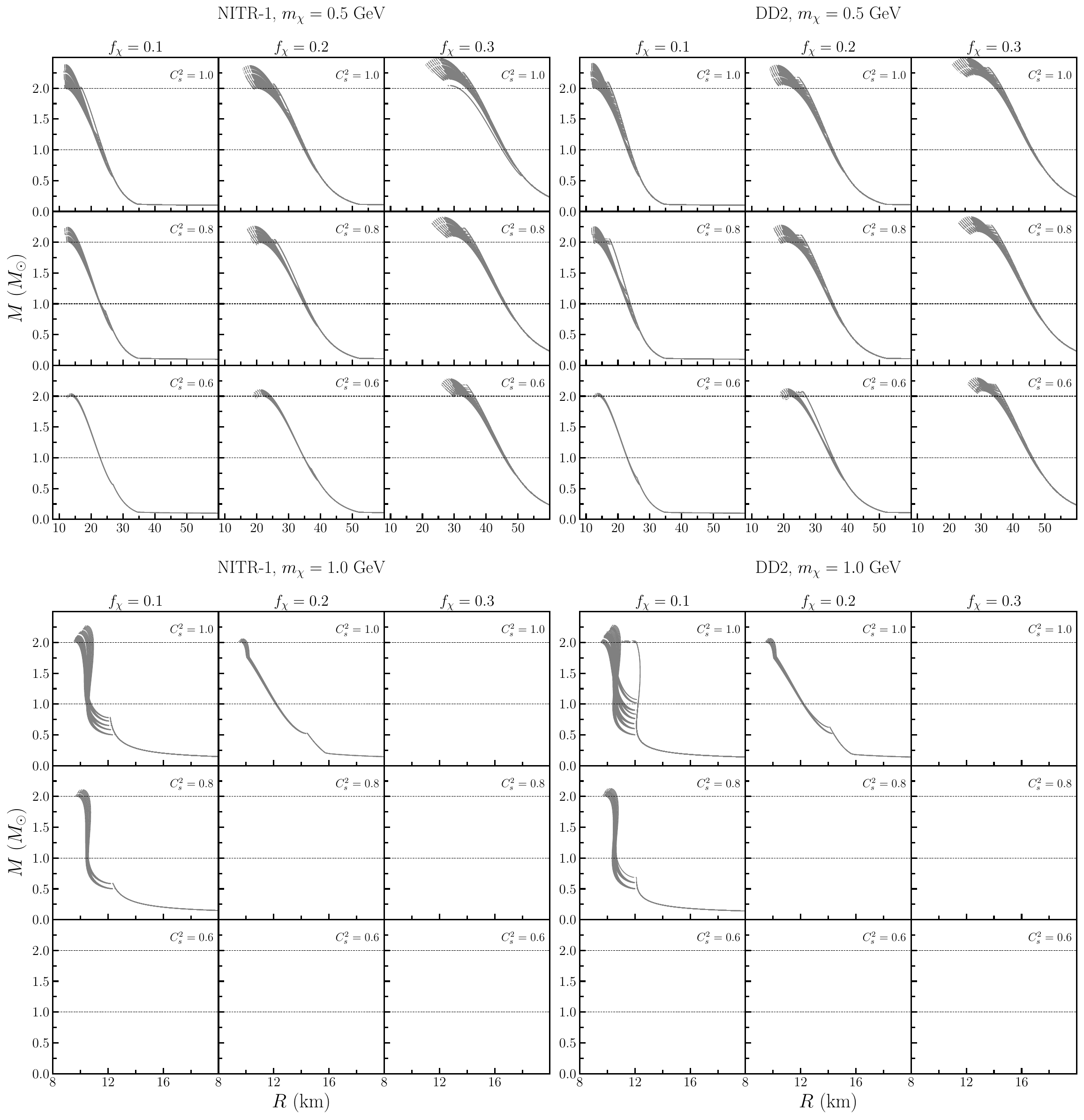}
		\caption{$M-R$ relations for hybrid star are shown for both NITR-1 and DD2 with $C_s^2 =0.6, 0.8,$ and 1.0 with three DMF having $m_\chi=0.5$ GeV (above) and 1.0 GeV (below).}
		\label{fig:mr_2f_hybrid}
	\end{figure*}
	\begin{figure*}
		\centering
		\includegraphics[width=0.5\textwidth]{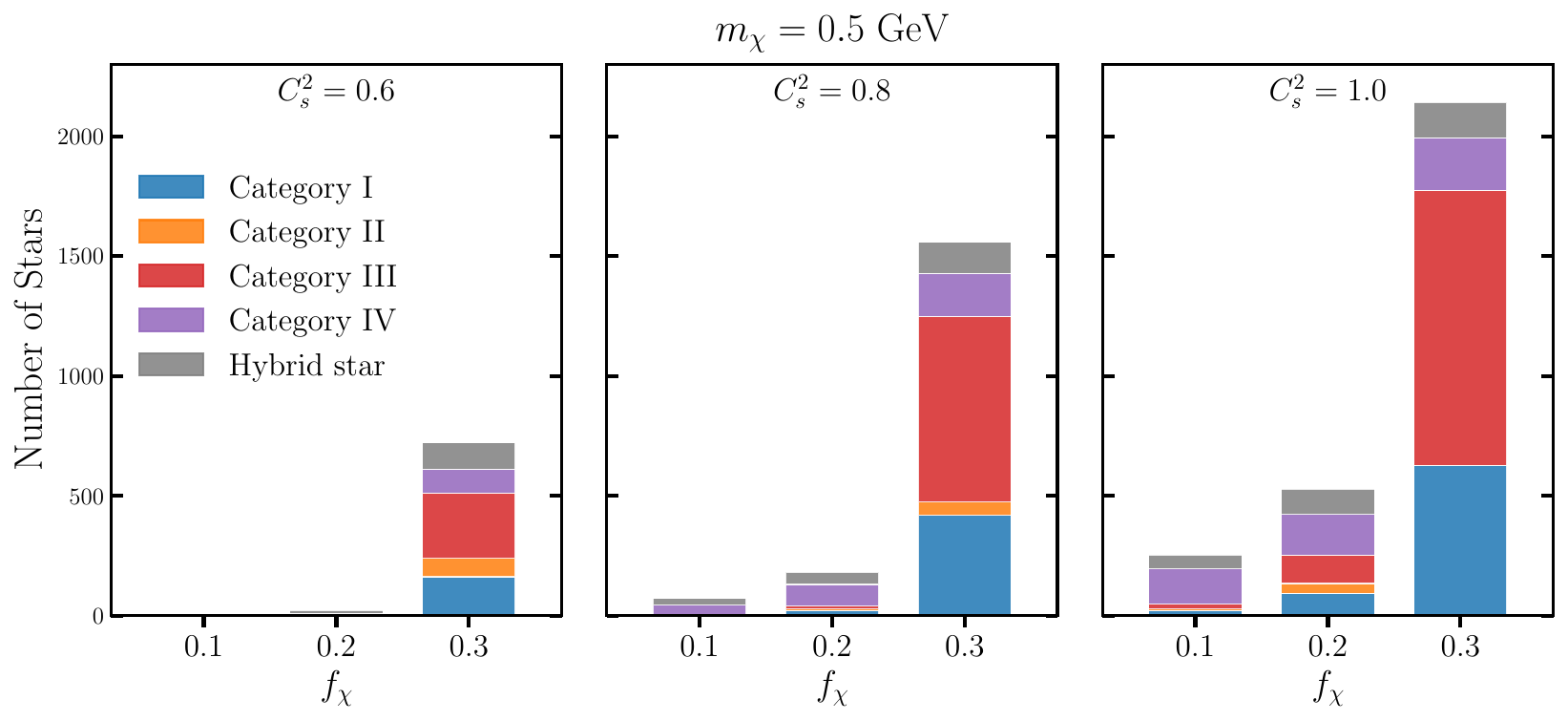}
		\includegraphics[width=0.5\textwidth]{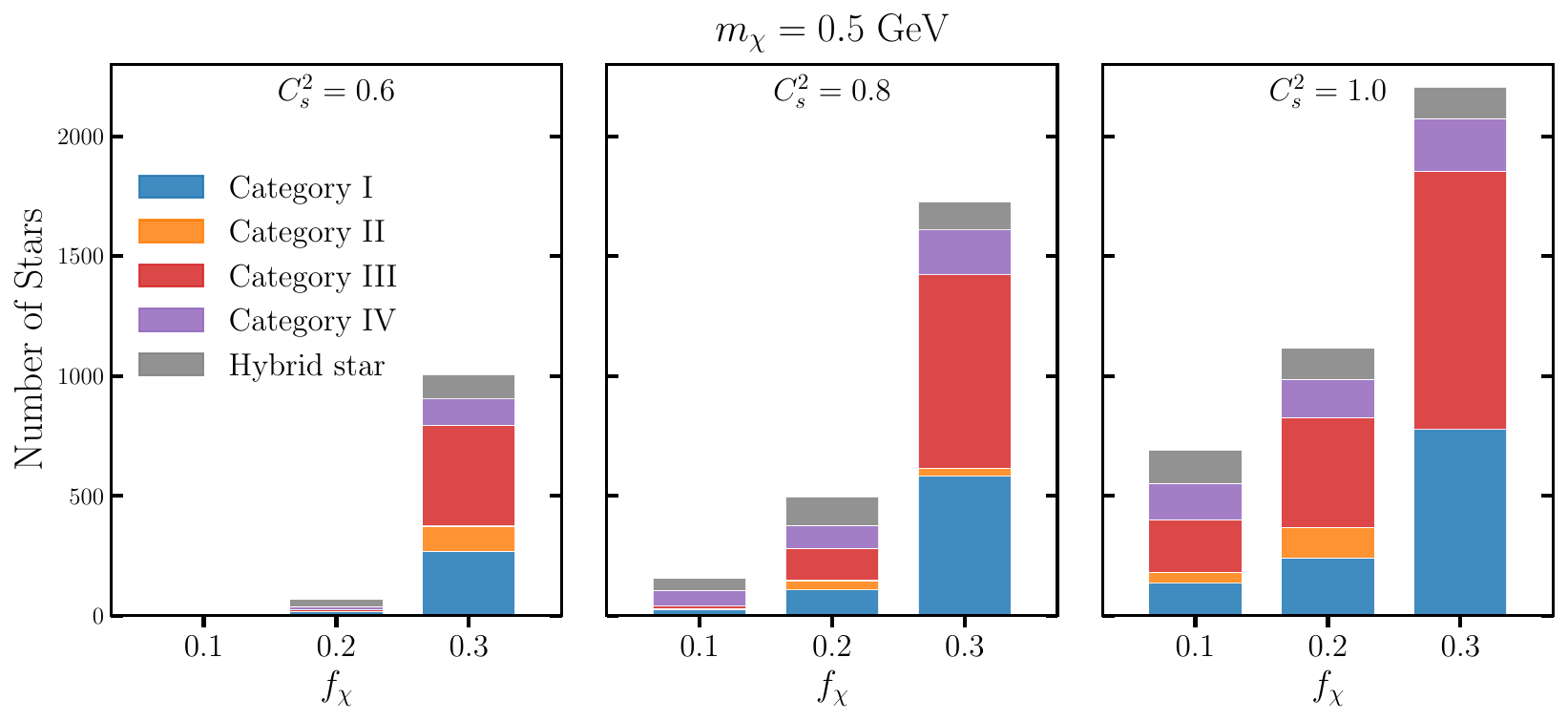}
		\caption{Number of twin star (colored by category) and hybrid star (gray) configurations as a function of DM fraction $f_\chi$ for NITR-1 (left) and DD2 (right), at $m_\chi = 0.5$~GeV, for $C_s^2 = 0.6$, $0.8$, and $1.0$.}
		\label{fig:stack}
	\end{figure*}
	\subsection{$M-R$ relations for twin star in two-fluid system}
	\label{subsec:2f_tss}
	The mass-radius ($M$-$R$) relations of the two-fluid DM-admixed twin star are shown in Fig.~\ref{fig:mr_2f_twin}, for NITR-1 (left) and DD2 (right), with $f_\chi = 0.1$, $0.2$, $0.3$ and $C_s^2 = 0.6$, $0.8$, $1.0$, at $m_\chi = 0.5$~GeV (upper panels) and $1.0$~GeV (lower panels). The color coding marks the four twin star categories. As discussed above, the halo or core character of the DM distribution is controlled by $m_\chi$ and governs all the trends described here. The upper panels reach radii as large as $\sim 50$~km, set by the extended DM halo, while the lower panels stay compact with $R \lesssim 12$~km.
	
	For the lighter particle ($m_\chi = 0.5$~GeV, halo regime), all four twin star categories appear, and increasing the DM fraction systematically enriches the twin star population. Since the halo only mildly compresses the baryonic core, adding more DM deepens the gravitational binding of the system and raises the central density of the baryonic component. The hadronic matter then crosses the hadron-quark transition more easily, quark cores form over a wider range of stellar masses, and more twin star branches are produced. This enrichment is strongest for the stiffest quark matter, $C_s^2 = 1.0$, because a twin star requires a stable disconnected high-density branch that only a sufficiently stiff quark phase can support. As the quark matter softens ($C_s^2 = 0.8$ and $0.6$), fewer twin star categories survive, setting up a direct competition between the softening of the quark phase, which works against twin star, and the DM compression, which works in their favor. In the softest case, $C_s^2 = 0.6$ with $f_\chi = 0.1$, no twin stars are found for either model, whereas raising the DM fraction to $f_\chi = 0.2$ and $0.3$ allows the DM compression to compensate for the soft quark phase and twin stars reappear.
	
	For the heavier particle ($m_\chi = 1.0$~GeV, core regime), the behavior is reversed. The compact DM core directly compresses the baryonic matter, lowers the maximum mass, and brings the star to gravitational instability at a lower central density. As the DM fraction increases, this instability is reached before the baryonic component can build up a quark core large enough to support a stable disconnected branch, so twin stars are strongly suppressed. They survive only at the smallest DM fraction with stiff quark matter ($f_\chi = 0.1$, $C_s^2 \gtrsim 0.8$), and essentially none remain at $f_\chi = 0.3$ for any $C_s^2$. The two nuclear models follow the same overall trend, with only a mild difference: the stiffer DD2 support twin star branches over a slightly wider region of parameter space than NITR-1. 
	\subsection{$M-R$ relations for hybrid star in two-fluid system}
	\label{subsec:2f_hss}
	The $M$-$R$ relations of the two-fluid DM-admixed hybrid stars are shown in  Fig.~\ref{fig:mr_2f_hybrid}, for the same models, DM masses, fractions, and sound speeds as in Fig.~\ref{fig:mr_2f_twin}. Each gray curve is a hybrid star sequence, that is, a compact star whose hadronic outer layers enclose a quark matter core. A hybrid star forms only when the central density crosses the hadron-quark transition, so the presence and extent of these sequences directly measure how easily a quark core can form and be stably supported. As in Fig.~\ref{fig:mr_2f_twin}, the halo regime ($m_\chi = 0.5$~GeV) produces extended sequences reaching $\sim 50$~km, while the core regime ($m_\chi = 1.0$~GeV) stays compact with $R \lesssim 13$~km.
	
	Hybrid stars follow the same competition established above for twin stars. In the halo regime ($m_\chi = 0.5$~GeV), hybrid sequences appear across almost the entire parameter space, and increasing $f_\chi$ progressively broadens the family of hybrid sequences. This effect is most pronounced for the stiffest quark matter. In the softest case, $C_s^2 = 0.6$ and $f_\chi = 0.1$, the hybrid branches reduce with certain number of sequences for both NITR-1 and DD2, but recovers a full family once the DM fraction is raised. In the core regime ($m_\chi = 1.0$~GeV), hybrid sequences are suppressed by increasing $f_\chi$ and vanish entirely at $f_\chi = 0.3$ for every $C_s^2$, surviving at $f_\chi = 0.1$ and a few with $f=0.2$ with stiff quark matter. As with the twin stars, the stiffer DD2 sustains hybrid sequences over a slightly wider region of parameter space than NITR-1.

	\subsection{Number of twin and hybrid stars}
	\label{subsec:number}
	Figure~\ref{fig:stack} shows the number of twin and hybrid star configurations as a function of $f_\chi$ for both NITR-1 and DD2, at $m_\chi = 0.5$~GeV. The most immediate observation is that the total population grows steeply with $f_\chi$ across all sound speeds and both nuclear models, confirming that DM in the halo regime actively increases the formation of exotic compact star configurations rather than suppressing them.
	
	Category III dominates the population at all values of $C_s^2$ and $f_\chi$. This is physically reasonable: Category III twin stars require only the quark branch to exceed $2\,M_\odot$, a condition that is more easily satisfied than the Category I requirement that both branches exceed this threshold. Category I configurations are present but always subdominant, and Category II remains consistently small across the entire parameter space. Category IV, which requires the hadronic maximum mass to fall below $1\,M_\odot$, appears only at large $f_\chi$, where the DM halo has sufficiently redistributed the stellar mass to push the hadronic branch below this limit.
	
	The sound speed has a clear and systematic effect. At $C_s^2 = 0.6$, the total population at $f_\chi = 0.1$ and $0.2$ is nearly zero for both models, and only at $f_\chi = 0.3$ does a significant population emerge. At $C_s^2 = 0.8$ and $1.0$, the population is already non-negligible at $f_\chi = 0.2$ and grows rapidly toward $f_\chi = 0.3$, reaching over 2000 configurations at $C_s^2 = 1.0$ and $f_\chi = 0.3$. This confirms that stiff quark matter and large DM fraction act cooperatively: each independently favors twin star formation, and together they produce the largest populations.
	
	Comparing the two nuclear models, DD2 consistently produces a larger total population than NITR-1 at fixed $C_s^2$ and $f_\chi$, reflecting its stiffer nuclear EOS. The difference is most visible at intermediate values, $C_s^2 = 0.8$ and $f_\chi = 0.2$, where DD2 already shows a substantial Category I and Category III population while NITR-1 remains sparse. At 
	$f_\chi = 0.3$, both models converge toward similarly large counts, suggesting that at high DM fraction the DM halo dominates over the nuclear stiffness in determining the twin star population.
	\begin{figure*}
		\centering
		\includegraphics[width=1.02\textwidth]{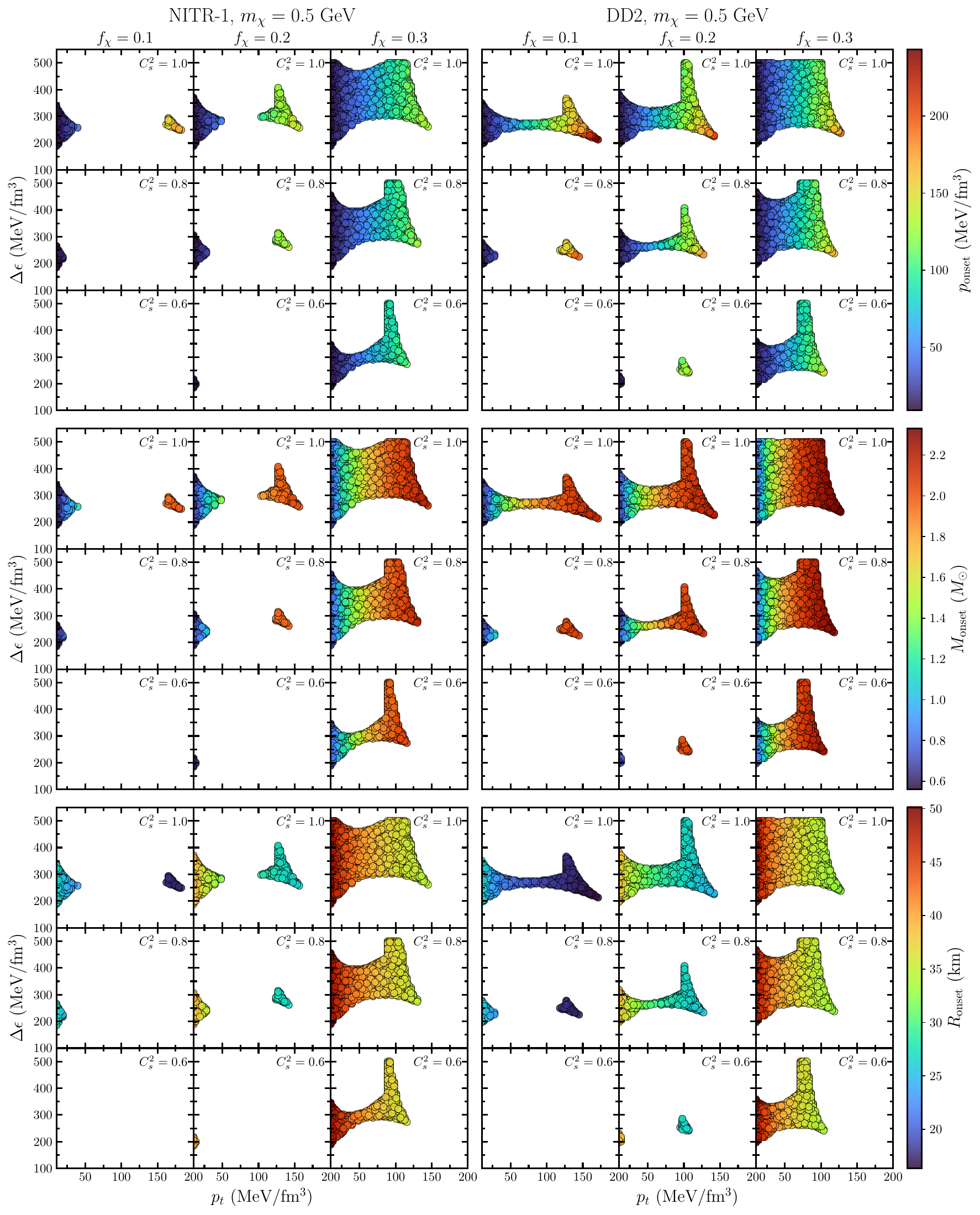}
		\caption{Onset properties of quark matter for two-fluid DM-admixed stars with $m_\chi = 0.5$~GeV. Each point is a twin-star configuration, shown in the plane of $p_t$ and $\Delta\epsilon$. Within each tile, columns give the DM mass fraction ($f_\chi = 0.1,\,0.2,\,0.3$) and rows the quark speed of sound ($C_s^2 = 1.0,\,0.8,\,0.6$). Points are colored by the onset pressure $p_{\rm onset}$ (above), mass $M_{\rm onset}$ (middle), onset radius $R_{\rm onset}$ (lower) for the central pressure, stellar mass, and radius at which the quark core first appears for the hadronic models NITR-1 and DD2.}
		\label{fig:onset}
	\end{figure*}
	\subsection{DM effects on the quark matter transition}
	\label{subsec:onset}
	Figure~\ref{fig:onset} is in many ways the central result of this work. It shows where the quark matter first appears inside a DM-admixed star, and what kind of star it appears in. The two axes describe the hadron-quark phase transition: $p_t$ is the pressure at which the transition takes place, and $\Delta\epsilon$ is the jump in energy density across it. Every point in the figure represents one successful twin star configuration, and its color gives the onset property, either the central 
	pressure ($p_{\rm onset}$), mass ($M_{\rm onset}$), or radius ($R_{\rm onset}$) of the star at the moment its quark core first forms. As before, the panels are arranged by sound speed ($C_s^2 = 1.0$, $0.8$, $0.6$ from top to bottom) and DM fraction ($f_\chi = 0.1$, $0.2$, $0.3$ from left to right), and all results shown here are for $m_\chi = 0.5$~GeV, that is, the halo regime.
	
	The first thing to notice is how much of the $p_t$-$\Delta\epsilon$ plane is filled with points. For stiff quark matter ($C_s^2 = 1.0$) and large DM fraction ($f_\chi = 0.3$), the points cover a broad region, meaning that a wide range of transition parameters can produce a quark core. As the DM fraction decreases or the quark matter softens, this region shrinks quickly, until at $C_s^2 = 0.6$ and $f_\chi = 0.1$ almost no valid configuration survives. This is the same trend already seen in the $M$-$R$ diagrams and in the twin star counts: more DM and stiffer quark matter open up the parameter space, while less DM and softer quark matter close it down.
	
	The $p_{\rm onset}$ panels confirm the most direct picture: the onset pressure simply follows $p_t$, growing from a few tens of $\mathrm{MeV\,fm^{-3}}$ on the left to a couple hundred on the right. The vertical axis, $\Delta \epsilon$, plays a smaller role in fixing the onset pressure, as the color bands are nearly vertical, but it controls how sharp the transition is, and therefore the configuration becomes a genuine twin star.
	
	The most important message comes from the color gradient in the $M_{\rm onset}$ panels. The color changes mainly with the transition pressure $p_t$. When $p_t$ is small, the quark core appears in light stars with onset masses as low as $\sim 0.6\,M_\odot$. When $p_t$ is large, the quark core appears only in heavy stars with onset masses close to $\sim 2\,M_\odot$. The reason is straightforward: $p_t$ is the pressure needed to start the transition, so a low $p_t$ is reached even near the center of a low-mass star, whereas a high $p_t$ is reached only in the dense core of a massive star. In other words, $p_t$ directly controls whether quark matter sets in early, in light stars, or late, only in the heaviest stars.
	
	The $R_{\rm onset}$ panels show the same story from the other side. A low $p_t$ corresponds to large onset radii (up to $\sim 45$-$50$~km), and a high $p_t$ to small onset radii ($\sim 20$~km). This mirror behavior is a direct consequence of the halo regime. For $m_\chi = 0.5$~GeV, light stars are physically large because their size is set by the extended DM halo, while heavy stars are compact. Since a low $p_t$ makes the quark core appear in a light star, it also makes it appear in a large star, and a high $p_t$ does the opposite. Taken together, the $M_{\rm onset}$ and $R_{\rm onset}$ panels tell us that an early onset of quark matter happens in light, extended stars, while a late onset happens in heavy, compact ones.
	
	Comparing the two nuclear models, DD2 fills a noticeably larger area of the $p_t$-$\Delta\epsilon$ plane than NITR-1 and reaches slightly higher onset masses and pressures, because its stiffer hadronic matter supports quark cores over a wider range of transition parameters. Overall, Fig.~\ref{fig:onset} provides the quantitative basis for the statements made earlier: the transition pressure decides when and in what kind of star quark matter appears, while the DM fraction and quark stiffness decide how much of the parameter space is available at all.
	\section{Summary and Conclusion}
	\label{summary}
	In this work, we have carried out a systematic study of DM-admixed hybrid and twin stars within a two-fluid framework, in which the baryonic (nucleon + quark) component and the DM component interact only through gravity. The hadronic sector was described by two relativistic mean-field EOSs of different stiffness, NITR-1 and DD2; the quark sector by the CSS parameterization with transition pressure $p_t$, energy density jump $\Delta\epsilon$, and speed of sound $C_s^2$; and the DM sector by a self-interacting fermionic model. Solving the two-fluid TOV equations over this parameter space, we classified the resulting configurations into hybrid stars and four twin star categories, counted their populations, and mapped the onset properties of the quark phase.
	
	Our central result is that DM influences the hybrid and twin star population through two distinct and opposite mechanisms, selected by the DM particle mass. For a light particle ($m_\chi = 0.5$~GeV), the DM forms an extended halo around the baryonic star. The halo adds gravitational binding from outside without itself occupying the core, gently raising the central pressure of the baryonic component and helping the star to reach the hadron-quark transition. Increasing the DM fraction deepens this effect, enlarging the accessible $(p_t, \Delta\epsilon)$ region and increasing the number of hybrid and twin stars. For a heavy particle ($m_\chi = 1.0$~GeV), the DM instead forms a compact core that sits at the stellar center and competes with the baryonic matter for the high-density region. Although this strongly compresses the baryonic component, it also drives the configuration to gravitational instability at a lower central density, so the star can no longer sustain a stable quark core. Increasing the DM fraction then suppresses hybrid and twin stars until they vanish entirely at $f_\chi = 0.3$. Summed over the full grid, the heavy DM core case yields nearly two orders of magnitude fewer twin stars than the light DM halo case.
	
	This reframes the role of DM in a way that is not a trivial consequence of the model construction. The often quoted statement that DM simply reduces the number of hybrid or twin stars holds only in the core regime; in the halo regime the opposite is true, and additional DM enhances the population. The DM particle mass therefore fixes the sign of the effect by determining whether the DM helps the star reach the quark phase or prevents it from sustaining one, while the DM fraction fixes its magnitude. The competition emphasized by the structure of the two-fluid framework is thus resolved in a definite and regime-dependent way, and this resolution, rather than the mere mapping of parameter space, is the physical 
	content of the present study.
	
	The onset analysis (Fig.~\ref{fig:onset}) makes this mechanism quantitative. The transition pressure $p_t$ is found to be the important variable: a low $p_t$ is reached even near the center of a light star, so the quark core appears early in low-mass and spatially extended configurations in the halo regime; a high $p_t$ is reached only in the dense cores of massive stars, so the onset is late and confined to compact, near-maximum-mass configurations. The onset mass and onset radius consequently vary in a mirrored fashion across the $(p_t, \Delta\epsilon)$ plane, while $\Delta\epsilon$ mainly controls whether the resulting branch is a connected hybrid star or a disconnected twin. When the $2\,M_\odot$ constraint is imposed, only configurations whose onset occurs early enough for the hybrid or twin branch to reach two solar masses survive, which further restricts the viable region of parameter space in both the DM core and DM halo scenarios.
	
	We have also assessed the robustness of these conclusions. The two nuclear EOSs, NITR-1 and DD2, yield the same qualitative picture; the stiffer DD2 merely widens the parameter windows slightly and supports twin branches over a marginally larger region. The findings are therefore not tied to a single hadronic EOS. The present study employs a self-interacting fermionic DM model. The halo--core dichotomy is a generic feature of any DM model in which the pressure support decreases with increasing particle mass, so the qualitative sign-reversal of the DM effect on the hybrid/twin star population is expected to persist for other fermionic DM descriptions. For bosonic DM candidates, however, Bose-Einstein condensation can substantially soften the DM equation of state, potentially shifting the halo--core boundary to different mass scales and modifying the quantitative thresholds found here. A systematic comparison across DM models, including bosonic and portal-coupled scenarios, is an important direction for future work. The classification scheme of Sec.~\ref{sec:claasification} was constructed at the stiffest value $C_s^2 = 1$ to allow direct comparison with previous work~\cite{Christian_2018, Montana_2019}, but the two-fluid results span $C_s^2 = 0.6$ to $1.0$, and the soft $C_s^2$ behavior is an integral part of the competition discussed above. Extending the analysis toward the perturbative QCD bound $C_s^2 = 1/3$ would only tighten the windows further. The two DM masses were chosen to bracket the halo and core regimes; a denser 
	grid in $m_\chi$ and $f_\chi$, and in particular intermediate masses near the halo-core boundary, would sharpen the quantitative boundaries but is not expected to change the qualitative conclusions.
	
	Several natural extensions follow from this study. The same framework can be applied to the tidal deformability, $f$-mode oscillations, and cooling of DM-admixed hybrid and twin stars, all of which are sensitive to the onset of quark matter and could provide observational discriminators between the core and halo regimes. A broader set of nuclear and quark EOSs, together with a finer DM grid, would allow the windows identified here to be confronted directly with current and forthcoming gravitational wave and NICER measurements. We leave these investigations for future work.
	\section*{Acknowledgments}
	We would like to thank J. E. Christian, G. Montana, H.-J. Schulze, and Vishal Parmar for discussions regarding CSS parameterization and Seidov's criterion and characterization conditions for twin star. Part of this computation was performed using the CINECA cluster. I gratefully acknowledge support from the Alexander von Humboldt Foundation through a Humboldt Research Fellowship.
	\section*{Data Availability}
	There are no publicly available research data or software supporting this manuscript. Requests for further information or data should be sent to the authors.
	\bibliography{twin star}
	\bibliographystyle{apsrev4-2}
\end{document}